\documentclass[pra,showpacs,twocolumn,superscriptaddress,aps]{revtex4-1}
\usepackage[pdftex]{graphicx}  
\usepackage{graphicx, xcolor}
\usepackage{dcolumn}
\usepackage{bm}
\usepackage{amsmath,amsthm,amssymb,bbold}
\usepackage{color}
\usepackage{verbatim}
\usepackage{nicefrac}
\usepackage{float}

\definecolor{darkGreen}{RGB}{0,110,0}
\definecolor{darkBlue}{RGB}{0,0,130}
\usepackage[colorlinks,citecolor=darkGreen,linkcolor=darkBlue,urlcolor=blue,hyperindex]{hyperref}

\begin{document}
\title{Experimental probes of Stark many-body localization}

\author{S.R.\ Taylor}
\affiliation{The Abdus Salam International Center for Theoretical Physics, Strada Costiera 11, 34151 Trieste, Italy}
\author{M.\ Schulz}
\affiliation{The Abdus Salam International Center for Theoretical Physics, Strada Costiera 11, 34151 Trieste, Italy}
\author{F.\ Pollmann}
\affiliation{Physics Department, Technical University of Munich, James-Franck-Stra{\ss}e 1, 85748 Garching, Germany}
\affiliation{Munich Center for Quantum Science and Technology (MCQST), Schellingstra{\ss}e 4, D-80799 M{\"u}nchen, Germany}
\author{R.\ Moessner}
\affiliation{Max Planck Institute for the Physics of Complex Systems, N{\"o}thnitzer Stra{\ss}e 38, 01187 Dresden, Germany}

\begin{abstract}

Recent work has focused on exploring many-body localization (MBL) in systems without quenched disorder: one such proposal is Stark MBL in which small perturbations to a strong linear potential yield localization.
However, as with conventional MBL, it is challenging to experimentally distinguish between non-interacting localization and true MBL.
In this paper we show that several existing experimental probes, designed specifically to differentiate between these scenarios, work similarly in the Stark MBL setting.
In particular we show that a modified spin-echo response (DEER) shows clear signs of a power-law decay for Stark MBL while quickly saturating for disorder-free Wannier-Stark localization.
Further, we observe the characteristic logarithmic-in-time spreading of quantum mutual information in the Stark MBL regime, and an absence of spreading in a non-interacting Stark-localized system.
We also show that there are no significant differences in several existing MBL measures for a system consisting of softcore bosons with repulsive on-site interactions.
Lastly we discuss why curvature or small disorder are needed for an accurate reproduction of MBL phenomenology, and how this may be illustrated in experiment. This also connects with recent progress on Hilbert space fragmentation in ``fractonic'' models with conserved dipole moment, and we suggest this as an auspicious platform for experimental investigations of these phenomena.
\end{abstract}

\maketitle

\section{Introduction}

The phenomenon of disorder-driven localization in quantum systems has inspired vast amounts of research, initially on Anderson localization in non-interacting systems \cite{Anderson1958,anderson1978local,abrahams1979scaling,Evers2008Review} and more recently on many-body localization (MBL) in interacting systems \cite{Basko2006Metal,Gornyi2005Interacting,Oganesyan2007Localization,Pal2010Many,Znidaric2008many,Bardarson2012Unbounded,Serbyn2013Universal,Bauer2013Area,Serbyn2013Local,DeLuca2013Ergodicity,BarLev2014Dynamics,Imbrie2016,Luitz2015Many,Singh2016Signatures}.
The MBL phase is a robustly non-thermalizing phase, and as such has implications for our understanding of how the breakdown of thermodynamics emerges from quantum-mechanical systems, as well as relevance to quantum computation \cite{Laumann2015Quantum}.
The phenomenology of the MBL phase, including the absence of transport \cite{Znidaric2008many,Pal2010Many}, Poissonian level statistics \cite{Oganesyan2007Localization}, area-law entanglement in eigenstates \cite{Bauer2013Area,Serbyn2013Local}, and logarithmic-in-time growth of entanglement \cite{Znidaric2008many,Bardarson2012Unbounded,Serbyn2013Universal}, can be explained in the language of local integrals of motion (LIOMs): an extensive set of mutually-commuting quasi-local operators that commute with the Hamiltonian, which emerge at strong disorder \cite{huse2013phenomenology,ros2015integrals,imbrie2017review}.

\begin{figure}
    \centering
    \includegraphics[width=\columnwidth]{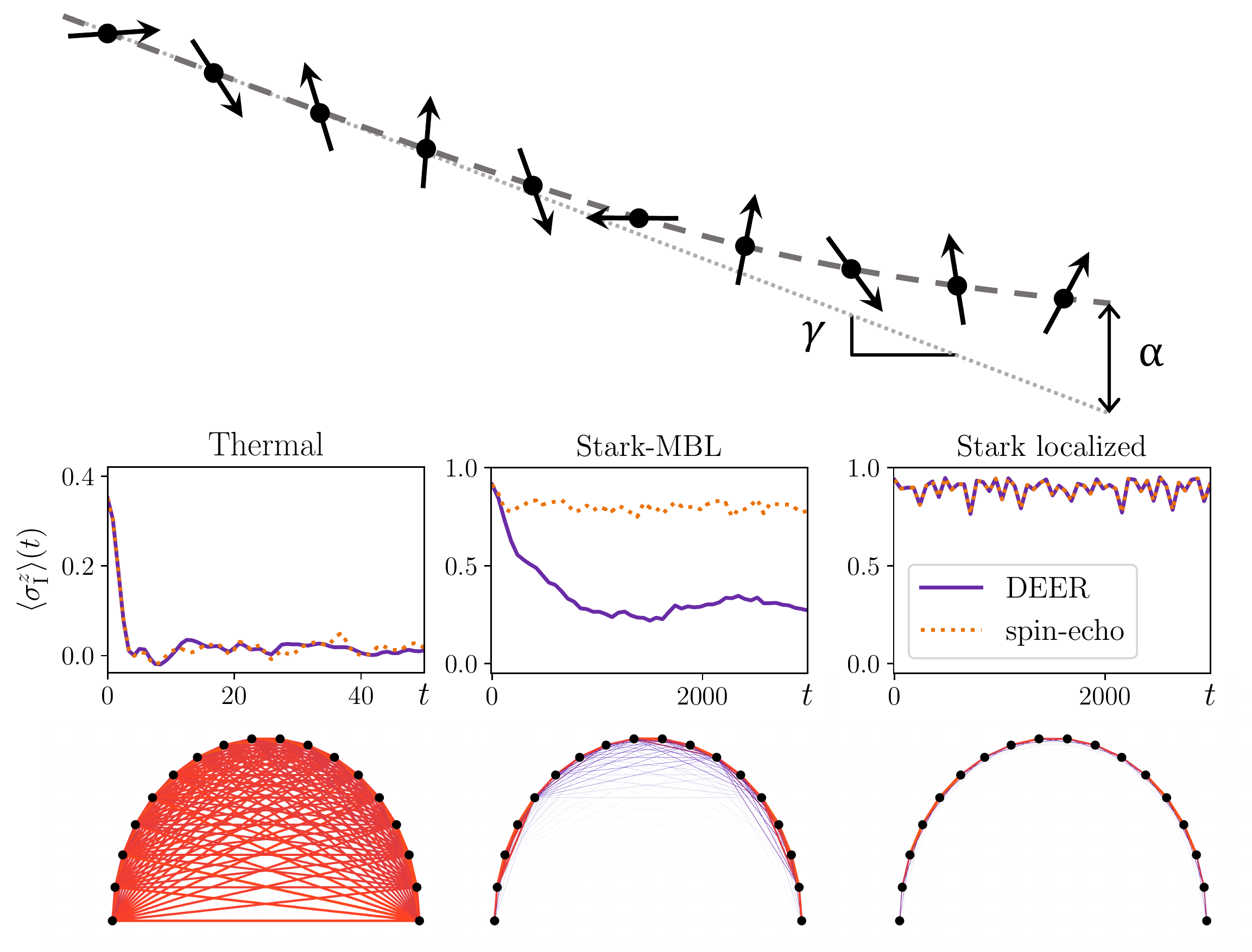}
    \caption{Top panel: Sketch of the system with a linear field ($\propto \gamma$) and quadratic contribution  ($\propto \alpha$). Middle panel: Characteristic behavior of the spin-echo and DEER responses for systems in the thermal (left), interacting Stark MBL (middle), and non-interacting Wannier-Stark (right) regimes. Bottom panel: Characteristic behavior of the eigenstate QMI in the same regimes, where the magnitude of the normalized QMI between two sites is indicated by the thickness of the lines between them and their color (red being strong and blue being weak).}
    \label{fig:Comparison}
\end{figure}

However, disorder is not the only way of achieving localized single-particle eigenstates. They can also be found in systems with a strong linear potential which goes by the name of Wannier-Stark localization \cite{Wannier1962Dynamics}. Alongside many other proposals for disorder-free localization \cite{Grover2014Quantum,Schiulaz2014Ideal,Yao2016Quasi,VanHorssen2015Dynamics,Hickey2016Signatures,DeRoeck2014Asymptotic,smith2017disorder,Brenes2018Many,Mondaini2017Many,Choudhury2018Frustration,Mamaev2019Quantum,Kuno2019Flat}, it was recently shown that many features of MBL can be observed in a system with a strong linear potential and interactions \cite{Schulz2019Stark,vanNieuwenburg2019From,ribeiro2019many,wu2019bath,Bhakuni2019Entanglement} (although some weak breaking of perfect linearity, for example through weak disorder or curvature, is required for a true comparison to conventional MBL).

From an experimental point of view there have been several promising attempts to study conventional MBL, including trapped ions \cite{Smith2016Many-body}, cold-atoms trapped in optical lattices \cite{Schreiber2015Observation,choi2016exploring,Lukin2019Probing}, and superconducting qubits \cite{Chiaro2019Growth}. Due to the short lifetimes of these experiments, a major challenge is to distinguish between non-interacting localization, interacting localization, or merely slow dynamics \cite{Panda2020Can}.
While theoretically the most striking difference between non-interacting and interacting localization is perhaps the slow growth of entanglement following a quench, in practice it remains challenging to measure due to its non-local nature \cite{Islam2015Measuring,Lukin2019Probing}.
As such, there have been proposals for more accessible measures that rely only on local measurements \cite{Roy2015Probing,Iemini2016Signatures,Serbyn2014Interferometric,Tomasi2017Quantum,Smith2016Many-body}.

In this work we explore how some of these proposals for experimental probes of MBL behave when applied to a Stark MBL system, particularly their ability to distinguish between thermalization and both interacting and non-interacting localization. We consider the Hamiltonian in the spin language:
\begin{equation}
    H = \sum_{n = 0}^{L-2} \left( s^x_n s^x_{n+1} + s^y_n s^y_{n+1} + \Delta s^z_n s^z_{n+1}\right) +  \sum_{n = 0}^{L-1} h_n s^z_n,
    \label{eq:H}
\end{equation}
where $s^{\mu}_n$ are spin-\nicefrac{1}{2} operators on site $n$ and $h_n = - \gamma n + \alpha n^2 / (L-1)^2$.
In this parametrization, $\gamma$ gives the strength of the uniform potential, and $\alpha$ is a small parameter that breaks the perfect uniformity of the potential.
A diagram of this system is shown in the top panel of Fig.~\ref{fig:Comparison}.
We measure time in the inverse of the nearest-neighbor exchange coupling, which we have set to 1.

We first explore the proposal of a modified non-local spin-echo protocol, akin to double electron-electron resonance (DEER), which can differentiate between thermal, MBL, and Anderson localized phases \cite{Serbyn2014Interferometric} (we note that these kinds of measurements have recently been used in an experiment on a conventional MBL system \cite{Chiaro2019Growth}). We find that this protocol has the same power in the Stark MBL system, which can be seen in the middle panel of Fig.~\ref{fig:Comparison}, where we show characteristic spin-echo and DEER responses for systems in the thermal, Stark MBL, and non-interacting Stark localized regimes.
We then study how quantum mutual information (QMI), which has been been shown to behave differently in thermal and localized systems \cite{Tomasi2017Quantum}, can be similarly applied to Stark MBL. The lower panel of Fig.~\ref{fig:Comparison} shows the behavior of the QMI in the eigenstates of the Hamiltonian \eqref{eq:H} in the three regimes; the black dots represent the sites of the system, and the magnitude of the QMI between them is indicated by the thickness and redness of the lines joining them.
Comparing Fig.~\ref{fig:Comparison} with the first figures of Refs.~\onlinecite{Serbyn2014Interferometric,Tomasi2017Quantum}, we see the similarities between conventional MBL and Stark MBL systems.
We also study the behavior of a bosonic system in a linear potential, given their prevalence in recent experiments, and find results similar to those in the fermionic system.
Finally we explore how the presence of curvature in the potential affects measurements of Stark MBL systems, in particular how curvature or small disorder are needed to localize pairwise hopping terms, and how this can potentially be discerned in experiment. We also show how Stark localization is intimately related to the recent discussion around fractons and Hilbert space fragmentation \cite{Pretko2017Subdimensional,Pai2019Localization, Sala2019Ergodicity,Khemani2019Local}.

\section{Spin echo measurements}

The recently proposed DEER protocol for MBL systems \cite{Serbyn2014Interferometric} allows one to probe the dynamical correlations between remote regions of a many-body system using only local manipulations and measurements.
In the context of conventional MBL, the DEER response is best illustrated by a description of the system in the LIOM representation \cite{Serbyn2013Local,huse2013phenomenology},
\begin{equation}
    \hat{H} = \sum_{i} \tilde{h}_i \tau_i^z + \sum_{ij}\mathcal{J}_{ij} \tau_i^z \tau_j^z + \sum_{ijk}\mathcal{J}_{ijk} \tau_i^z \tau_j^z \tau_k^z + \dots,
    \label{eq:H_lbit}
\end{equation}
where the LIOMs $\tau_i^z$ are effective spin-\nicefrac{1}{2} operators with eigenvalues $\pm 1$.
The couplings $\mathcal{J}_{ij},\mathcal{J}_{ijk},\dots $ fall off exponentially with separation, with a characteristic localization length $\xi$.
Starting from an arbitrary eigenstate (i.e. a product state in the LIOM basis), the protocol consists of applying a $\frac{\pi}{2}$-pulse to $\tau^{z}_{\text{I}}$, denoted by  $R^{\pi/2}_{\text{I}}$, such that $R^{\pi/2}_{\text{I}} \vert \uparrow \rangle_{\text{I}} = \left(\vert \uparrow \rangle_{\text{I}} + \vert \downarrow \rangle_{\text{I}} \right)/\sqrt{2}$.
After a time $t/2$ the precession of $\tau^{z}_{\text{I}}$ is reversed by applying a $\pi$-pulse, denoted by  $R^{\pi}_{\text{I}}$, and after a further time $t/2$ a final $R^{\pi/2}_{\text{I}}$ returns the LIOM to its initial configuration.
In concrete experiments, one would manipulate and measure the physical spins and not the LIOMs.
However, as was argued in Ref.~\onlinecite{Serbyn2014Interferometric}, this gives qualitatively similar results, and becomes quantitatively correct in the limit of short localization length.
When performed on a localized system (i.e. when \eqref{eq:H_lbit} is valid) the expectation value of $\tau^{z}_{\text{I}}$ at the end of this protocol is equal to its initial value.
However, in an extended system (where one cannot speak of LIOMs) the expectation value of $\sigma^{z}_{\text{I}}$ decays to zero as $t$ increases.

As with many experimental probes, this protocol does not distinguish between non-interacting and interacting localized phases.
The DEER protocol \cite{Serbyn2014Interferometric} introduces an additional perturbation (also a $\pi/2$-pulse) on a remote region II consisting of $N$ LIOMs at a distance $d \gtrsim \xi$ from $\tau^{z}_{\text{I}}$, applied at the same time as the $\pi$-pulse.
Assuming that the remaining spins are in a state with definite $\tau^z$, all interactions except those between spin I and region II are decoupled, and as such the DEER protocol directly measures the influence of region II on spin I.
More concisely, the many-body wave function under the DEER protocol takes the form
\begin{equation}
    \vert \psi (t) \rangle =  R^{\pi/2}_{\text{I}} e^{-i\hat{H}t/2}   R^{\pi}_{\text{I}} R^{\pi/2}_{\text{II}}  e^{-i\hat{H}t/2}     R^{\pi/2}_{\text{I}} \vert \psi(0)\rangle,
\end{equation}
where $R^{\pi/2}_{r} = \prod_{j\in r}\left(\hat{\mathbb{1}} -i \hat{\tau}^{y}_{j} \right)/\sqrt{2}$, and $R^{\pi}_{r} = \left(R^{\pi/2}_{r}\right)^{2}$.
The DEER response is defined as $\mathcal{D}(t) := \langle \psi(t)\vert \tau^z_{\text{I}} \vert \psi(t) \rangle$.
The dynamical phases accumulated due to interactions between regions I and II are different before and after the $\pi$-pulse, and therefore will not cancel, resulting in a decay of $\mathcal{D}(t)$. Contrastingly, in a non-interacting system the response does not decay and the behavior matches that of the original spin-echo protocol.
The differing behaviors in the three regimes can be seen in the middle panel of Fig.~\ref{fig:Comparison}.

\begin{figure}
    \centering
    \includegraphics[width=\columnwidth]{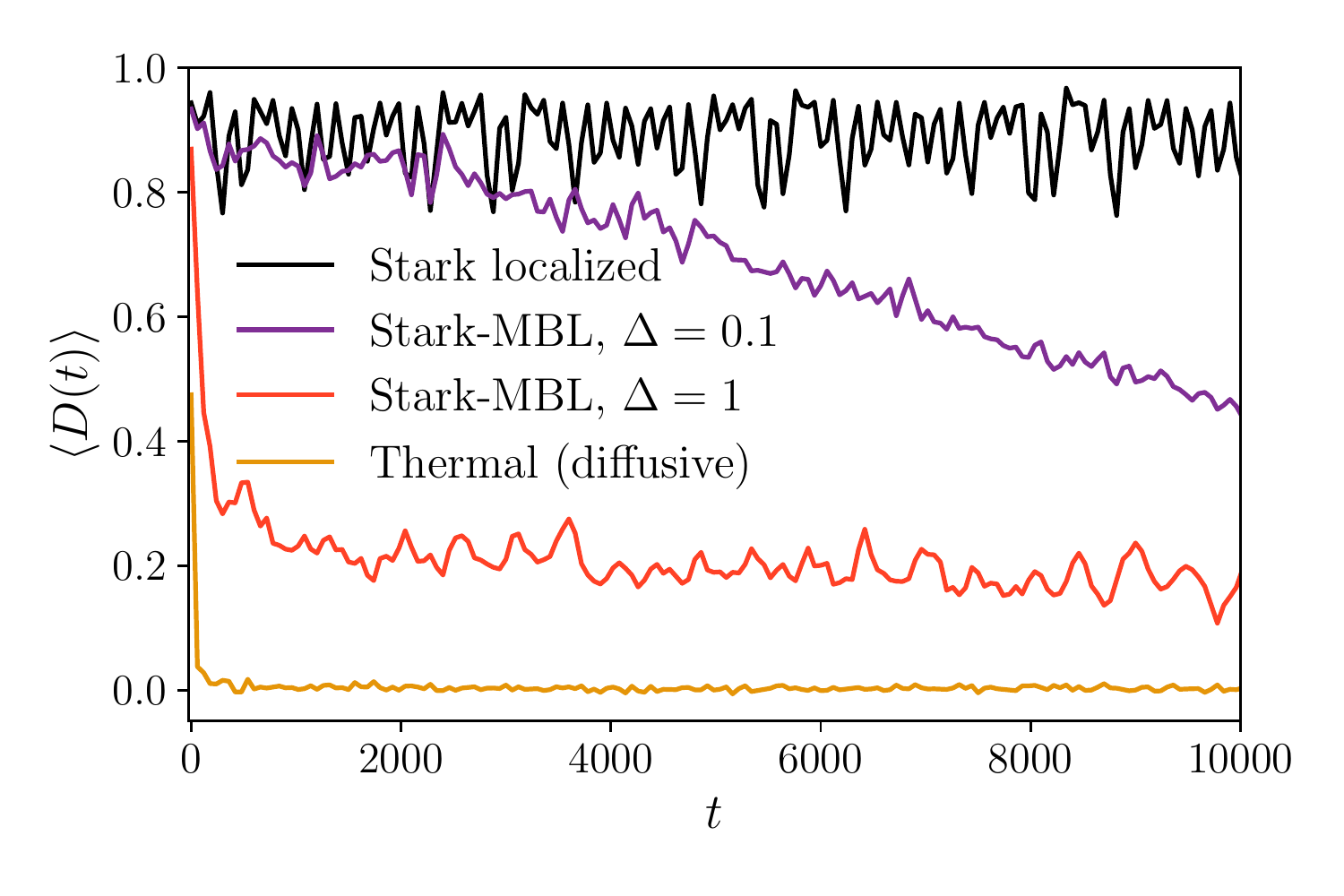}
    \caption{The short- to medium-time behavior of the thermally averaged DEER response ($ \langle \mathcal{D}(t)\rangle $) in the Stark MBL, non-interacting Stark, and thermalizing diffusive regimes. We see a clear difference between the three regimes, and observe (see the black Stark localization curve) that the DEER response is able to distinguish interacting and non-interacting localization, which a spin-echo protocol would not be able to do (see Fig.~\ref{fig:Comparison}). The parameters here are $L = 12$, $M = 50$ with a potential of $\gamma = 3$, $\alpha = 2$. The thermal system is exposed to uniformly-distributed on-site disorder with half-width 1 rather than a linear potential with curvature. For all cases $d = 3$ and $N = 7$.}
    \label{fig:DEER_exp}
\end{figure}

We now aim to show that this diagnostic for conventional MBL (bar the disorder averaging) holds in a similar fashion for a Stark MBL system.
We study the time evolution and response functions by exact diagonalization of \eqref{eq:H} on systems of $L = 12$ lattice sites with open boundary conditions. The DEER and spin-echo responses are calculated numerically as $\mathcal{D}(t) = \langle \psi(t)\vert \sigma^z_{\text{I}} \vert \psi(t) \rangle$. We consider mixed initial states by choosing $\vert \psi(0) \rangle$ as a random eigenstate satisfying $\mathcal{D}(0) > 0$, and by performing thermal averaging over $M$ states chosen randomly from the entire spectrum, denoted by the single brackets $\langle \mathcal{D}(t) \rangle$. We also verified that the experimentally simpler implementation of random product states (in the physical spin basis) yields the same result (see Appendix~\ref{app:DEER}).

The results for the DEER protocol are summarized in Fig.~\ref{fig:DEER_exp}.
Short- to intermediate-time simulations are most relevant, as experiments can access only up to $\sim 10^2-10^3$ hopping times \cite{choi2016exploring,Lukin2019Probing,Chiaro2019Growth}.
In Fig.~\ref{fig:DEER_exp} we show the DEER response for interacting and non-interacting Stark localization and the thermal diffusive regime. 
While the spin-echo response is either relatively constant or rapidly decaying to zero, the DEER protocol shows three different responses: a) it stays constant in the non-interacting Stark localized regime, b) it shows a slow decay for Stark MBL, and c) it rapidly decays to zero in the thermal regime.

While results at short to intermediate times are more relevant experimentally, we may also calculate the DEER response after thermal averaging for long times. In Appendix~\ref{app:DEER} we observe a power-law decay, albeit noisy, that spans several decades. We also observe that the saturated value for the DEER response, $ \langle \mathcal{D}(\infty)\rangle$ is nearly independent of interaction strength and region separation $d$, but only depends on the size of region II. We thus conclude that in the case of sufficient curvature the Stark MBL and conventional MBL systems behave qualitatively the same under the DEER protocol.

\section{Quantum mutual information}

Quantum entanglement has proven to be a powerful diagnostic in studies of many-body localization \cite{Abanin2019Colloquium}, particularly the growth of entanglement under unitary dynamics following a quench into a non-entangled state.
The appeal of this measure is that it distinguishes between single-particle localization (in which the entanglement entropy saturates after a short time) and many-body localization, which is characterized by unbounded logarithmic-in-time growth of the entanglement entropy \cite{Znidaric2008many,Bardarson2012Unbounded,Serbyn2013Universal}.
However, measuring the entanglement between macroscopically large regions, such as the half-chain entanglement that is often considered in theoretical studies, is very challenging experimentally.
Recent studies have turned to entanglement measures between small parts of the system \cite{Iemini2016Signatures,Tomasi2017Quantum}, with one example being the quantum mutual information (QMI) between two spatially separated sites $m$ and $n$:
\begin{equation}
    I(m,n) := S(\{ m \}) + S(\{ n \}) - S(\{ m, n \}),
\end{equation}
where $S(\sigma)$ is the von Neumann entanglement entropy between the set of sites in $\sigma$ and their complement.
This quantity has been shown to have different properties in the thermalizing and MBL phases, and it is an appealing measure because the two-site entanglement can be accessed experimentally \cite{Fukuhara2015Spatially,Jurcevic2014Quasiparticle}.

In the MBL phase, $I(m,n)$ decays exponentially with $|m-n|$ when evaluated in the eigenstates of the conventional MBL Hamiltonian, while in the thermal phase $I(m,n)$ has comparable values for all $|m-n|$ \cite{Tomasi2017Quantum}.
We observe the same behavior in Stark MBL systems as in conventional MBL, as can be seen in the lower panels of Fig.~\ref{fig:Comparison}, which show the QMI between sites of the system (indicated by the black dots).
The value of the QMI, relative to the largest value in the system, is indicated by the thickness of the lines and by their color, with thick red lines indicating relatively large values and thin blue lines indicating small values.
The values are averaged over 20 eigenstates from the middle of the spectrum.
In the Stark MBL regime the QMI decays with distance and in the non-interacting Wannier-Stark localized system there is no long-range QMI, just as for disorder-driven localization.
This indicates a clear difference in the behavior of the QMI of highly excited eigenstates in the various regimes of the Stark MBL Hamiltonian \eqref{eq:H}. As individual eigenstates cannot be resolved experimentally, in the following we will investigate the dynamical properties of the QMI after a global quench.

Following a quench starting from an initially non-entangled state, the QMI spreads logarithmically in time in the MBL phase, while in the thermal phase the spreading is algebraic in time \cite{Tomasi2017Quantum}.
It was shown that many-body and single-particle localization can be distinguished by the quantity
\begin{equation}
    X^2(t) = \sum_{n=1}^{L-1} n^2 \; \mathcal{I}_n(t) - \left( \sum_{n=1}^{L-1} n \; \mathcal{I}_n(t) \right)^2,
\end{equation}
where $\mathcal{I}_n := I(0,n)$.
Logarithmic-in-time growth is observed in the MBL phase, but a saturation to a constant value is seen in a non-interacting localized system \cite{Tomasi2017Quantum}.
We expect the QMI in the MBL phase to then decay at exponentially large times, which are not relevant for experimental purposes \cite{DeTomasi2019Efficiently}.
On the other hand, in a thermalizing system the QMI grows quickly at short times before decaying \cite{AlbaQuantum2019}.

\begin{figure}
    \centering
    \includegraphics[width=\columnwidth]{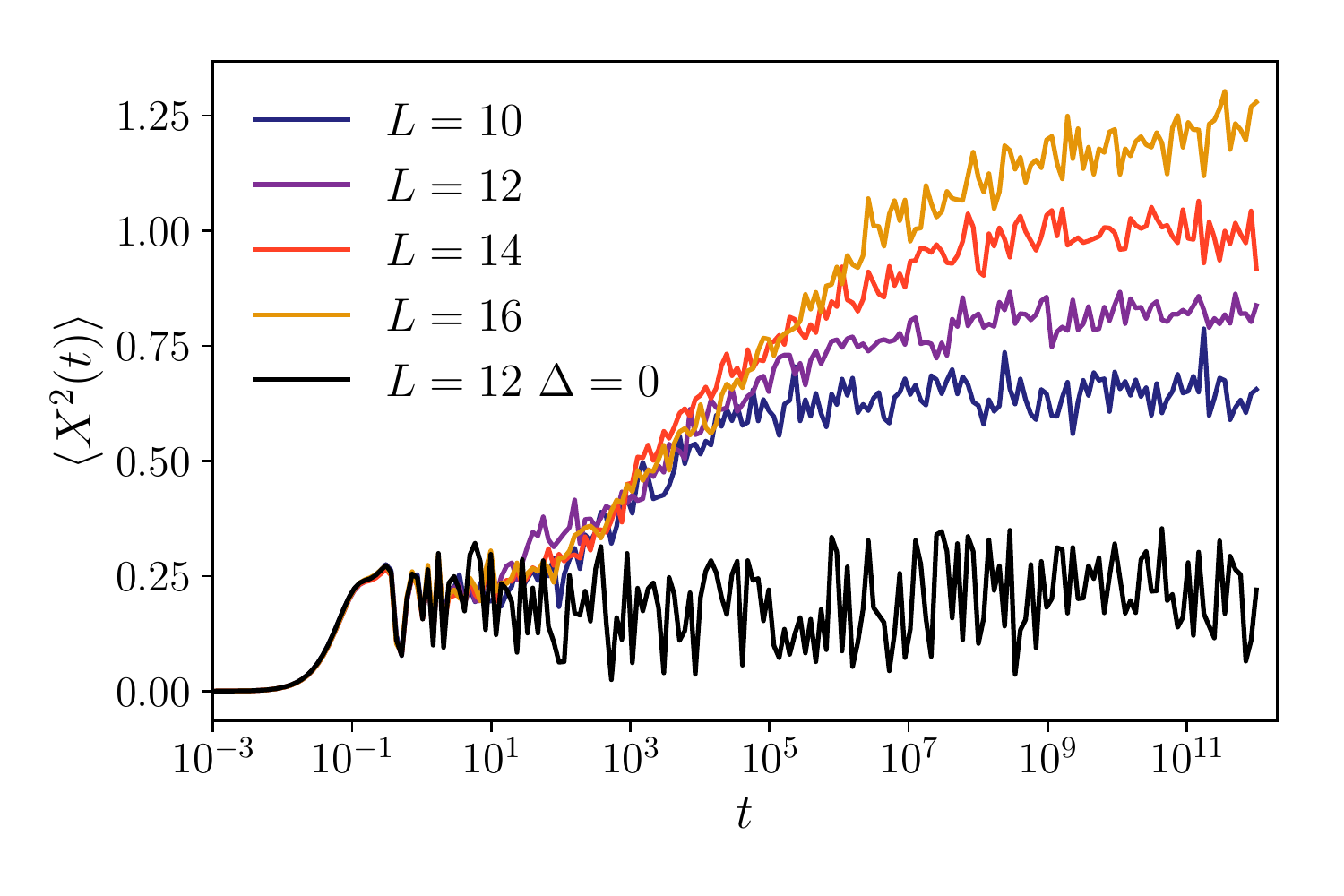}
    \caption{Logarithmic growth of $\langle X^2(t) \rangle$ for a system with $\gamma=2$, $\alpha=2$ and $\Delta=0.1$, with a non-interacting example for comparison. The results are averaged over at least 500 initial states, except for $L=10$ which is averaged over every Fock state in the magnetization sector.}
    \label{fig:QMI}
\end{figure}

Fig.~\ref{fig:QMI} shows the evolution of $\langle X^2(t) \rangle$ for several system sizes in the Stark MBL regime with weak interactions ($\gamma=2$, $\alpha=2$, $\Delta=0.1$), where the angled brackets indicate an average over initial conditions (Fock states in the physical basis).
The result from a non-interacting system is included for reference.
The rapid saturation in the non-interacting system is evident, while the logarithmic-in-time growth (before a finite-size saturation) is present in the interacting system as observed in conventional MBL.
Thus we see that the QMI is a good probe in both conventional MBL and Stark MBL systems, and that the phenomenologies of the two types of system are qualitatively similar.

\section{Stark MBL with interacting bosons}

Theoretical studies of MBL have mostly focused on systems of fermions, as the Hilbert space grows less severely with system size than for bosons due to the exclusion principle \cite{Sierant2018Many}, allowing numerical methods to better approach the thermodynamic limit.
However, a number of experimental studies with cold atoms have explored MBL in systems of interacting bosons.
A bosonic realization of the Stark MBL Hamiltonian \eqref{eq:H} has the form:
\begin{equation}
    H_B = \sum_{j=0}^{L-2} \left( b^{\dagger}_{j} b^{\mathstrut}_{j+1} + h.c. \right) + \sum_{j=0}^{L-1} \left( \frac{U}{2} n_j (n_j-1) + h_j n_j \right),
\end{equation}
where $b_j^{\dagger}$ ($b_j^{\mathstrut}$) are the creation (annihilation) operators for a boson on site $j$, $n_j := b_j^{\dagger}b_j^{\mathstrut}$, and $h_j$ has the same form as before.
We work with systems of $N_b = L/2$ bosons, in analogy with half-filling in fermionic systems, in order to reach larger system sizes.

\begin{figure}
    \centering
    \includegraphics[width=\columnwidth]{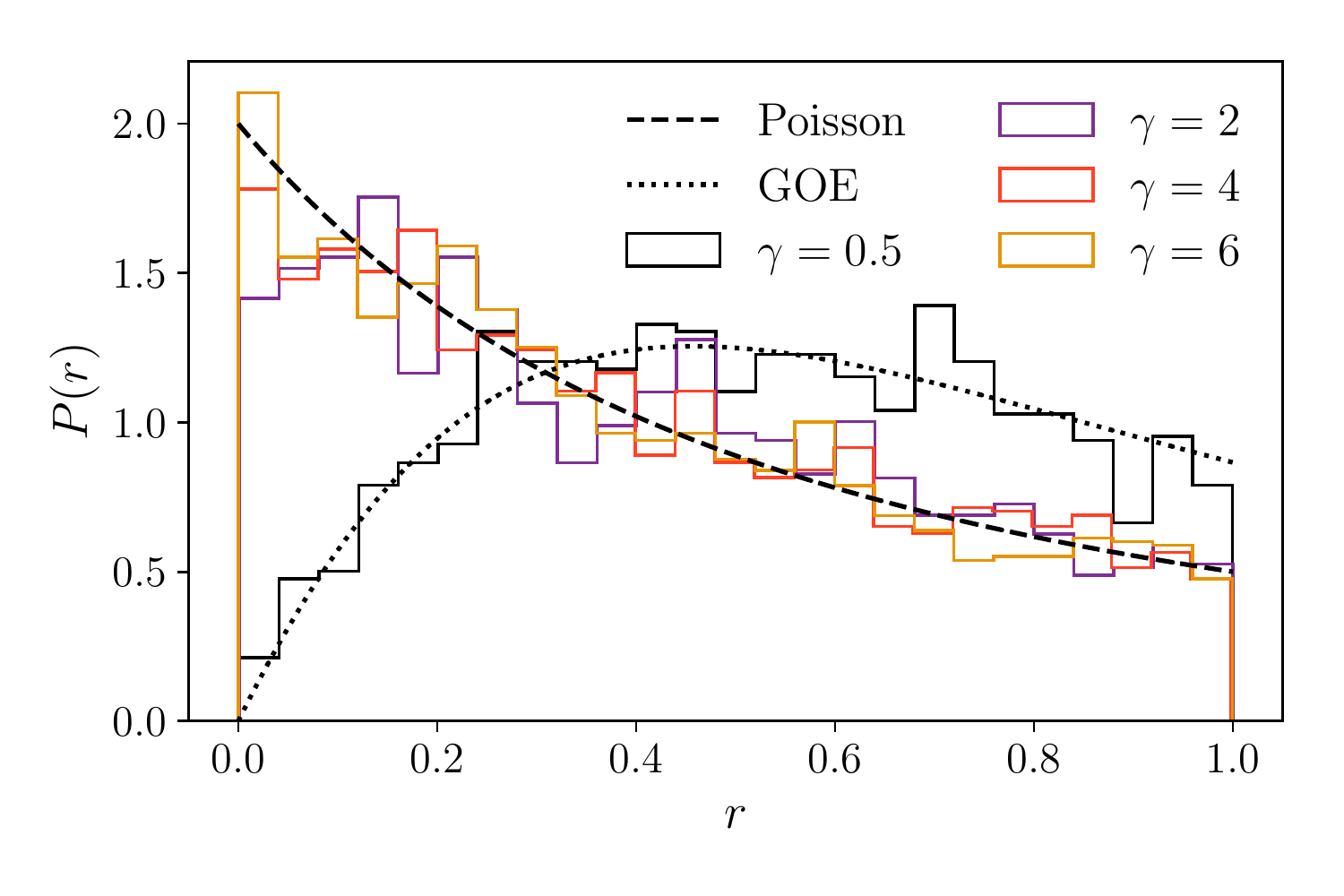}
    \includegraphics[width=\columnwidth]{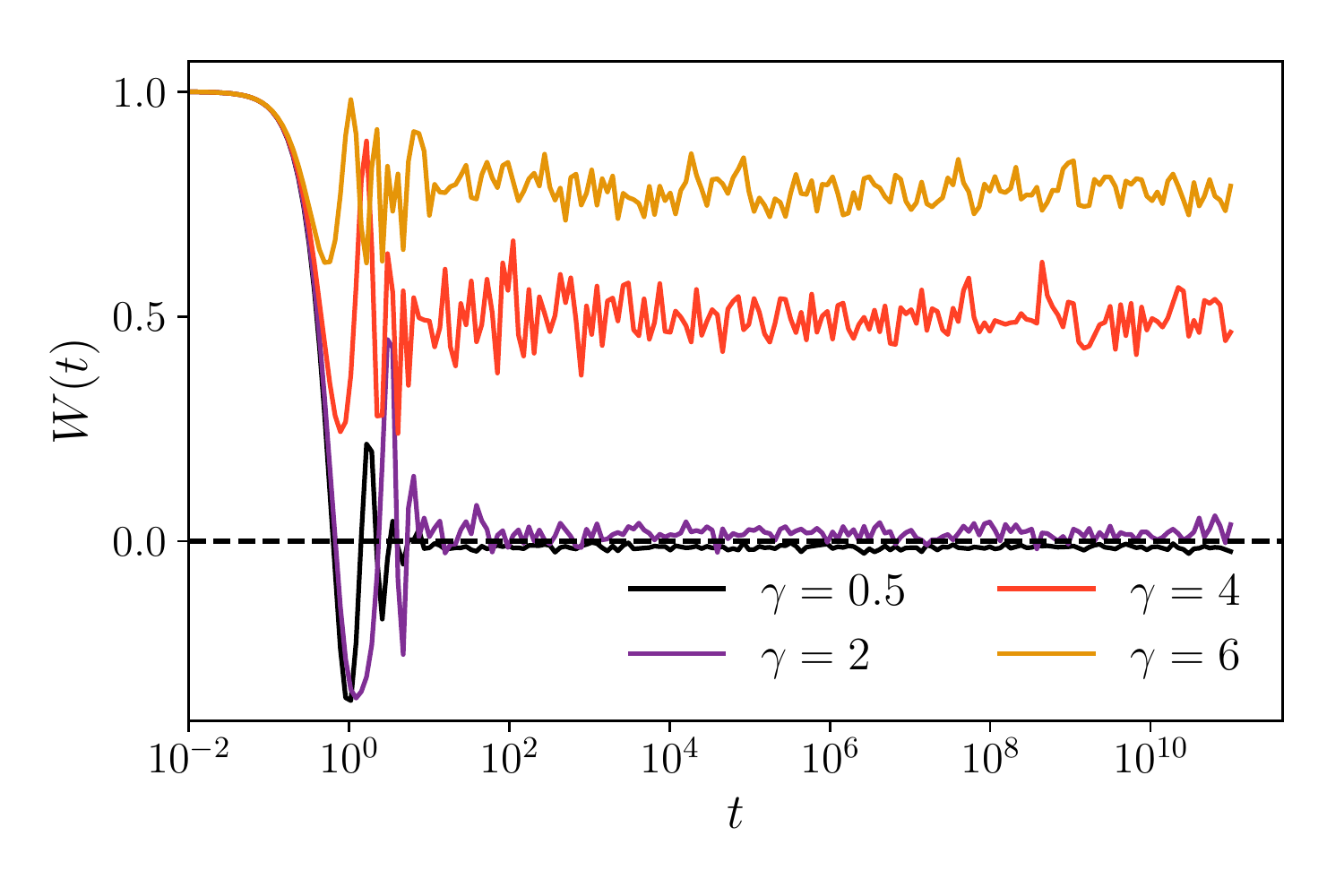}
    \caption{Top: the $r$ parameter for a system of interacting bosons. Bottom: decay of charge-density wave order. Both measures show signatures of Stark MBL. All data is for a system of $N_b=6$ bosons on a lattice with $L=12$ sites, with interaction strength $U=1$ and curvature $\alpha=2$.}
    \label{fig:Bosons}
\end{figure}

The $r$ parameter of the level statistics, defined as $r_n = \min(\delta_n, \delta_{n+1}) / \max(\delta_n, \delta_{n+1})$ with $\delta_n = E_{n+1}-E_{n}$ for the ordered energy levels $E_n$, is a standard diagnostic of MBL behavior.
In a thermalizing system, $r$ should exhibit random-matrix statistics (for this model the statistics of the Gaussian orthogonal ensemble (GOE)), while a localized system should obey Poisson statistics \cite{Atas2013Distribution}.
The upper panel of Fig.~\ref{fig:Bosons} shows histograms of $r$ for a range of potential gradients, indicating a change from thermalization at very small $\gamma$ to localization as $\gamma$ increases.

Another common test for MBL is the failure of the system to lose memory of an initially ordered state.
We initialize the system in a charge-density wave state, with one boson on every even site and every odd site empty, then observe whether this ordering is destroyed by unitary time evolution.
This measure is particularly significant as it has been applied to a number of experimental studies of MBL.
We measure the imbalance between odd and even sites with the quantity
\begin{equation}
    W (t) = \frac{\sum_{j=0}^{L-1} \; (-1)^j \; \langle n_j \rangle (t)}{\sum_{j=0}^{L-1} \; \langle n_j \rangle (t)},
\end{equation}
which is equal to 1 in the initial state, but will decay to zero if the populations of odd and even sites equalize.
The lower panel of Fig.~\ref{fig:Bosons} shows the decay of $W(t)$ as a function of time, showing that the order is still present at late times for large $\gamma$, consistent with the Stark MBL behavior observed in fermionic systems.

\section{The purely linear potential, and the effect of adding curvature}

A key feature of Stark MBL is that one cannot obtain true MBL phenomenology in a purely linear potential. This is perhaps most visible in the bipartite entanglement growth after a quench, which is much faster than logarithmic, an effect that was attributed to the many degeneracies in a purely linear potential \cite{Schulz2019Stark}. This is consistent with the level statistics, which shows neither Poisson nor Wigner-Dyson statistics, but instead a sharp peak at zero energy difference.
In this section we outline a two-particle perturbative argument for the origin of this behavior, the crux of which is the generation of an effective hopping scale much reduced from its bare value. This would explain how an otherwise surprisingly small amount of curvature or disorder can destroy the  behavior characteristic of the purely linear set-up.
We then argue that this fast growth of entanglement in the absence of curvature may be a signature of thermalization in large systems, but we cannot observe this with the finite system sizes that we can simulate.

\subsection{Two interacting particles in a linear potential}
\label{sec:twoparticles}

We consider two particles in a strong linear potential of slope $\gamma$, with a long-ranged interaction $V(s)$, and perturbatively introduce a nearest-neighbor hopping $J$.
We may parametrize the states by their center of mass (``dipole'') coordinate $m = (n_1 + n_2)/2$ and their relative coordinate $s = n_2 - n_1$, where $n_1$ and $n_2$ are the positions of the two particles.
The state $| m;s \rangle$ has energy $E = 2 \gamma m + V(s)$, and we note that in the absence of interactions the states for a fixed $m$ value form a degenerate subspace.
While a single hop is energetically suppressed due to the large energy difference between the initial and final states, pairwise hops in opposite directions (i.e. $| m;s \rangle \to | m;s \pm 2 \rangle$) result in a much smaller energy difference.
To lowest order, the effective amplitude for this hopping in the $s$ coordinate has the form:
\begin{equation}
\label{equ:pert_hop}
    \Tilde{J}\sim \frac{\vert J \vert^2}{\gamma - V^{\prime}a} + \frac{\vert J \vert^2}{-\gamma - V^{\prime}a} \sim \frac{\vert J \vert^2}{\gamma^2}V^{\prime}a,
\end{equation}
where $a$ is the lattice constant, so that $V^\prime a$ is the difference in interaction energy for pair configurations differing by one hop, with the two terms corresponding to the two possible intermediate states $| m\pm1/2; s\pm1 \rangle$.
We therefore see that the hopping in $s$ is small in $J / \gamma$, and it requires interactions to prevent the amplitudes for the two virtual processes cancelling.
For a short-ranged potential this hopping requires higher-order virtual processes: the effective pair-hopping amplitude vanishes until the particles have made a virtual excursion to within an interaction range of each other.
The effective pair hopping is thence exponentially suppressed with interparticle distance.

This weak hopping must compete with the diagonal matrix elements introduced by the interactions.
In systems with longer-ranged interactions (e.g.\ interactions that decay with distance algebraically), these diagonal matrix elements are likely sufficiently large that they dominate over the weak hopping and the system remains localized.
On the other hand, in systems with short-ranged interactions (e.g.\ nearest-neighbor or rapidly exponentially decaying ones) it is conceivable that the system delocalizes through these pairwise hopping processes. Settling this question therefore requires an analysis of the balance between diagonal and off-diagonal matrix elements at high order in perturbation theory, with the latter themselves varying with $s$. Appendix~\ref{app:ED2P} presents simulations of the  two-particle problem. There, it appears that localization persists even for the case of a purely linear potential. However, this requires   arbitrary precision numerics, which we implement using the mpmath package in Python \cite{mpmath}, as even the ``noise'' introduced by using only standard machine precision alters the numerical results qualitatively. Such a delicate phenomenon is likely very hard to access experimentally: observing delocalization of distant particles would require exceedingly long times (e.g.\ times exponentially large in $L$), well beyond the decoherence times of present experimental platforms even for moderate system sizes. However, this issue clearly warrants a more detailed investigation as a point of principle. 

In the case of delocalization, any curvature or weak disorder introduced to the onsite potentials would only have to compete with the much reduced effective hopping, and would be expected to easily cause localization in the $s$ coordinate. In the full many-body setting the reduced hopping must additionally navigate a much more complicated energy landscape; at any rate, our reasoning agrees qualitatively with the entanglement structure of the many-body eigenstates (discussed in the next section).

\subsection{The effect of curvature on eigenstate entanglement}

\begin{figure}
     \centering
    \includegraphics[width=0.48\textwidth]{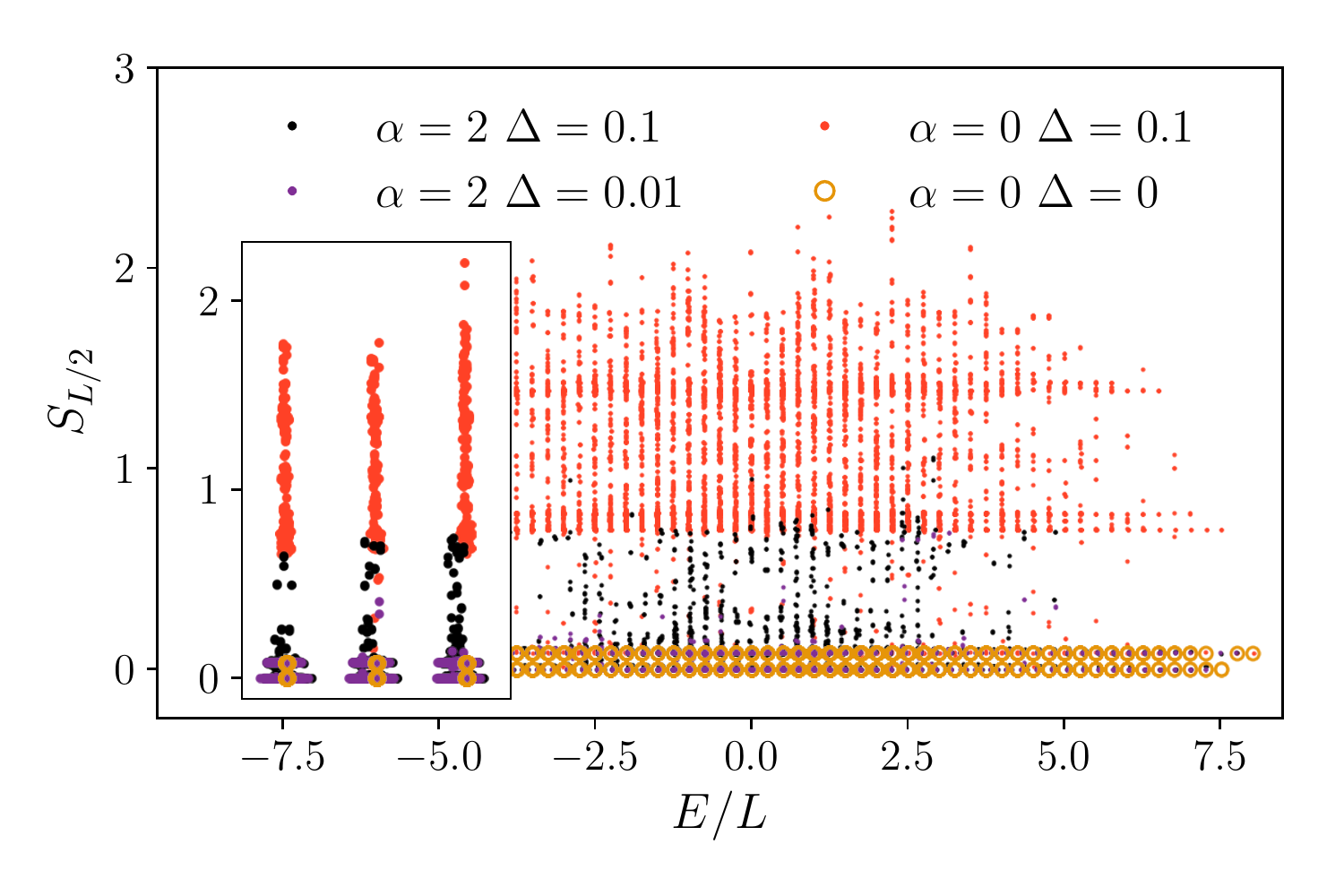}
    \includegraphics[width=0.48\textwidth]{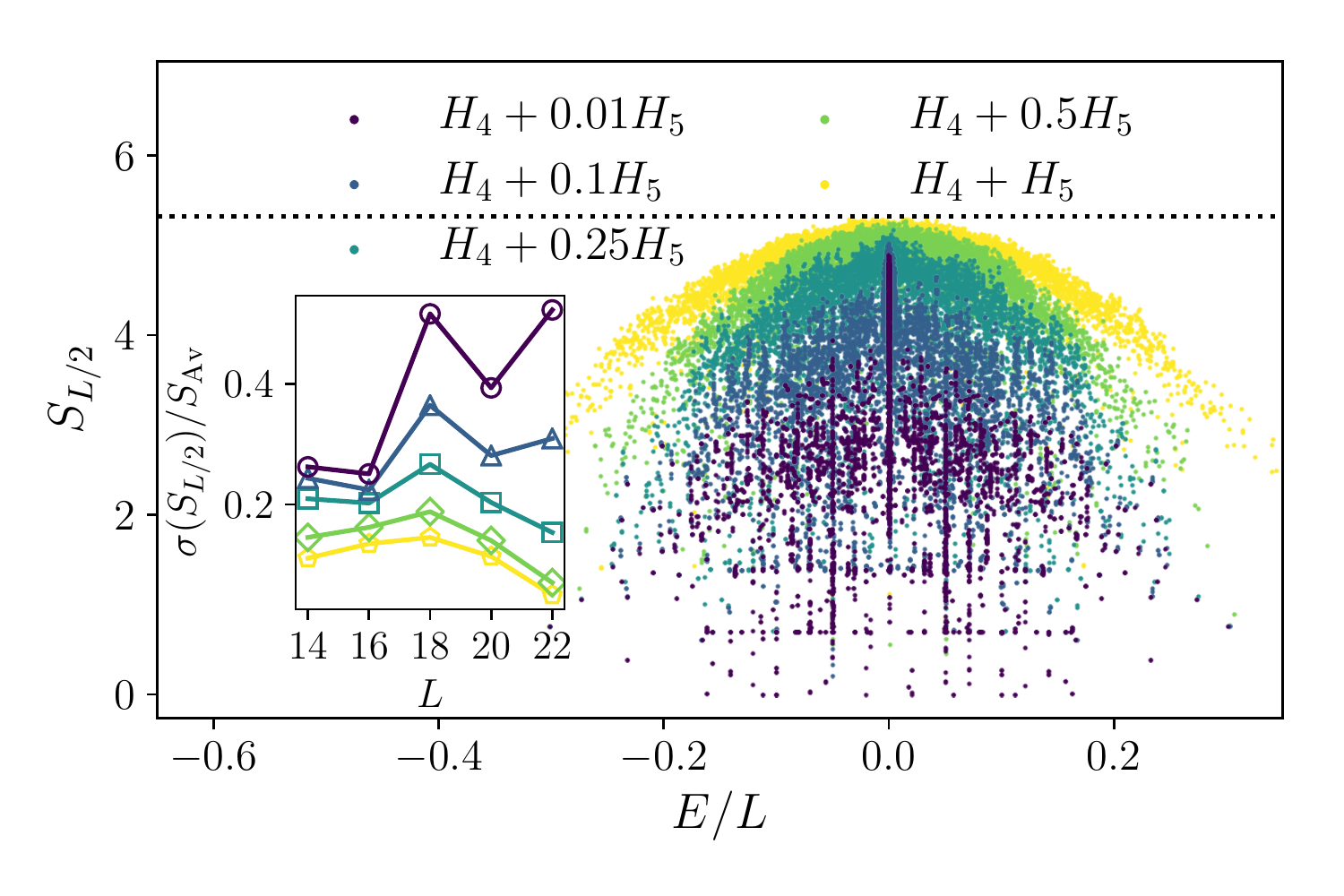}
     \caption{Upper panel: the half-chain entanglement $S_{L/2}$ as a function of energy for the eigenstates of a Stark localized XXZ system with $L=16$, $\gamma = 4$, and $\alpha=2$. Comparisons with the corresponding $\alpha = 0$ and non-interacting systems are shown.
     Lower panel: the half-chain entanglement in a dipole-conserving model with $L=20$ for different relative strengths of $H_4$ and $H_5$.
     The dotted line shows the average entanglement entropy of a random state, and the inset shows the size of the fluctuations of $S_{L/2}$ as a function of system size. The color scheme in the inset is the same as in the main panel.}
     \label{fig:S_E}
 \end{figure}

We probe whether or not the full many-body system thermalizes by examining whether the half-chain entanglement entropy of the system's eigenstates is a smooth function of eigenenergy, as demanded by the eigenstate thermalization hypothesis (ETH).
In the upper panel of Fig.~\ref{fig:S_E} we show the half-system entanglement entropy $S_{L/2}$ as a function of the eigenenergy for both interacting and non-interacting systems with and without curvature.
The clustering of eigenvalues seen in Fig.~\ref{fig:S_E} is a result of the Hamiltonian becoming approximately proportional to the dipole moment operator $P = \sum_n n s^z_n$ for strong $\gamma$; this is an analog in the spin language to the center of mass, and the clusters correspond to the center of mass subspaces described in the two-particle argument above.
The inset shows the same data, magnified around three clusters from the center of the spectrum.

For a non-interacting system with $\alpha=0$ (gold circles), the approximate center of mass subspaces are quite distinct, and the entanglement is very small.
We can understand this as a result of the eigenstates being combinations of (approximately exponentially) Stark-localized single-particle orbitals.
With the addition of interactions (red points), pair-hopping  within the appropriate center of mass subspaces (as described in Sec.~\ref{sec:twoparticles}) becomes possible, and there is a significant amount of entanglement. However, $S_{L/2}$ is not a smooth function of energy, indicating that this is not a signature of thermalization in the sense of the ETH.
In a system with both curvature and interactions (black and purple points), i.e. a Stark MBL system, we see that the entanglement entropy is always small, indicating that the eigenstates are similar to product states of localized single-particle orbitals, particularly in the limit of very weak interactions.
This is in keeping with the LIOM picture of conventional MBL, and is therefore consistent with the Serbyn-Papi\'{c}-Abanin argument for logarithmic entanglement growth \cite{Serbyn2013Universal}.
In the absence of curvature we cannot make this argument, and indeed the entanglement entropy was shown to grow much faster than logarithmically in Fig. 2 of Ref.~\onlinecite{Schulz2019Stark}.

\subsection{Relation to dipole-conserving ``fractonic'' models}

As mentioned above, for very large $\gamma$ the Hamiltonian \eqref{eq:H} becomes approximately proportional to $P$.
In this limit we may treat the hopping and interaction terms as perturbations to $H_0 = - \gamma P$ which, due to the large energy cost, are unable to cause transitions between states with different dipole moments.
One can then perturbatively construct an effective Hamiltonian for a single degenerate subspace with $P=P_0$, in which the off-diagonal couplings are generated by virtual transitions to states with $P \neq P_0$ \cite{Soliverez1969Effective}.
The physics of the Stark MBL system with $\alpha=0$ is therefore connected to the physics of models that conserve dipole moment, which exhibit a breakdown of thermalization due to Hilbert space fragmentation \cite{Sala2019Ergodicity,Khemani2019Local}.
A spin-\nicefrac{1}{2} realization of such a model could consist of two terms: $H_4 = \sum_n \left(s^+_{n} s^-_{n+1} s^-_{n+2} s^+_{n+3} + h.c. \right)$ and $H_5 = \sum_n \left(s^+_{n} s^-_{n+1} s^-_{n+3} s^+_{n+4} + h.c. \right)$, which are the leading-order terms appearing in the perturbative analysis described above.
It was demonstrated that a system composed of $H_4$ alone does not thermalize, but in the system $H_4 + H_5$ charge correlation functions were found to relax to their thermal values, with the exception of a small number of outlying eigenstates \cite{Sala2019Ergodicity}.
The dipole-conservation in these models results in restrictions on the mobility of excitations, such as is seen in the physics of fractons \cite{Pretko2017Subdimensional,Pai2019Localization}.
As one would expect terms with the forms of $H_4$ and $H_5$ to appear in a perturbative expansion of $H$ with $\alpha=0$, there appears to be some inconsistency between this picture and the lack of thermalization seen in the upper panel of Fig.~\ref{fig:S_E}.
However, we can reconcile the results by comparing the thermalization of dipole-conserving models with different relative weights of $H_4$ and $H_5$.
The lower panel of Fig.~\ref{fig:S_E} shows the half-chain entanglement entropy of eigenstates as a function of energy for a selection of different combinations of $H_4$ and $H_5$ for $L=20$ in the largest connected sector with $P=0$.
While for a system solely composed of $H_4$ and a system composed of $H_4+H_5$ we recover non-thermalization and thermalization respectively, as seen in Ref.~\onlinecite{Sala2019Ergodicity}, for systems where $H_5$ is present only weakly we see an apparent breakdown of thermalization.
In an expansion of $H$ one would expect higher order terms to appear with decreasing weights, so the non-thermalizing behavior in the two models is consistent.

To quantify to what extent the system thermalizes we have analyzed the fluctuations of the entanglement entropy by calculating the standard deviation of $S_{L/2}$ within a small energy window $E \in [ -0.025 L, 0.025 L]$, rescaled by the average, which decreases with increasing $L$ in a thermalizing system.
This is shown in the inset of the lower panel of Fig.~\ref{fig:S_E}, where we see a trend towards thermalization in large systems.
For $H_4 + 0.25 H_5$, which appears non-thermalizing in the main panel, the fluctuations start to decrease for $L \gtrsim 18$.
For $H_4 + 0.1 H_5$ we do not yet see a monotonic decrease with increasing system size, but when compared to $H_4 + 0.01 H_5$ the data appears consistent with thermalization at larger $L$.
This would suggest that both these dipole-conserving models with weak $H_5$ and also Stark MBL systems with $\alpha=0$ will thermalize if the system is large enough.

\begin{figure}
    \centering
    \includegraphics[width=\columnwidth]{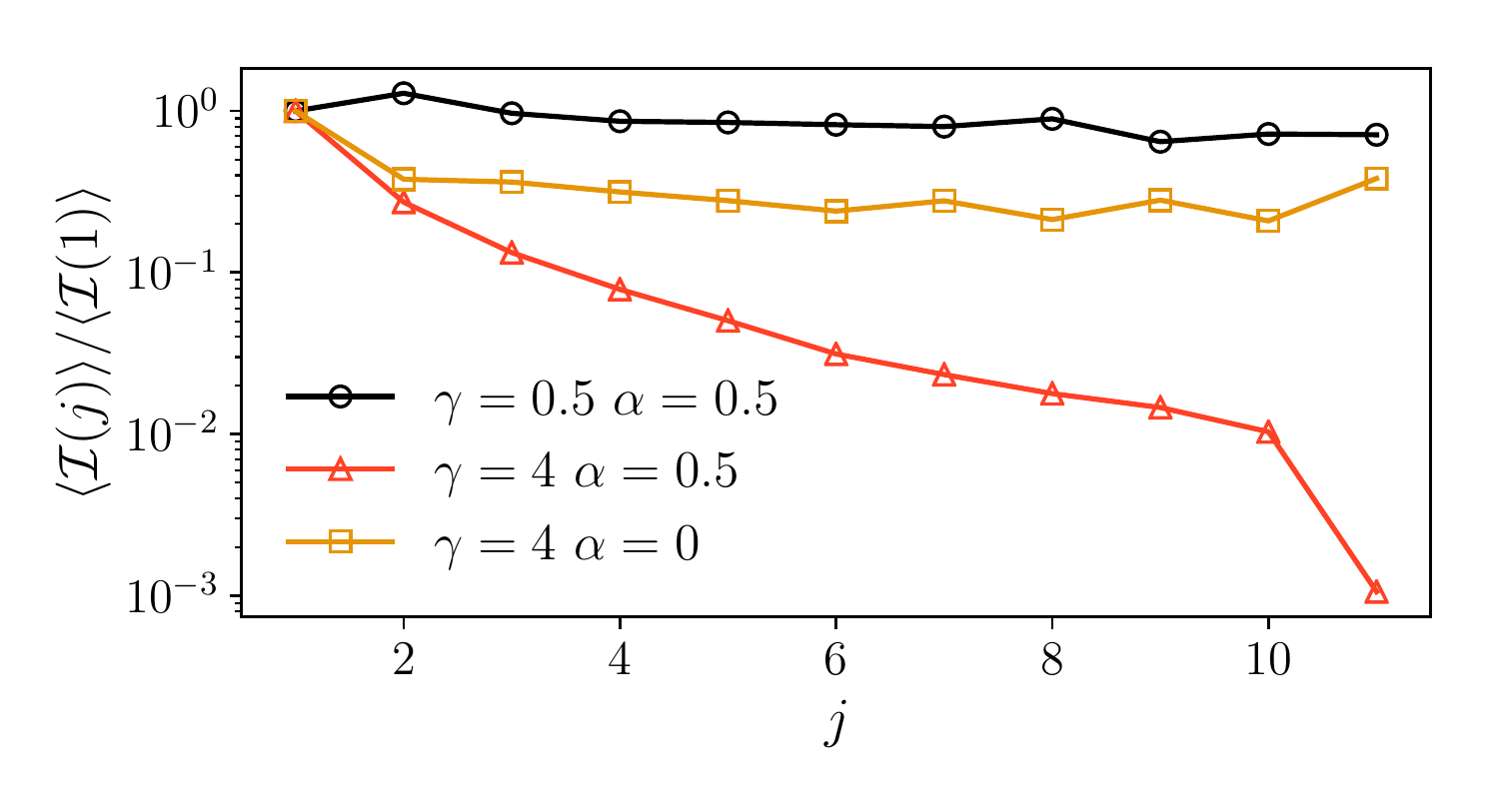}
    \includegraphics[width=\columnwidth]{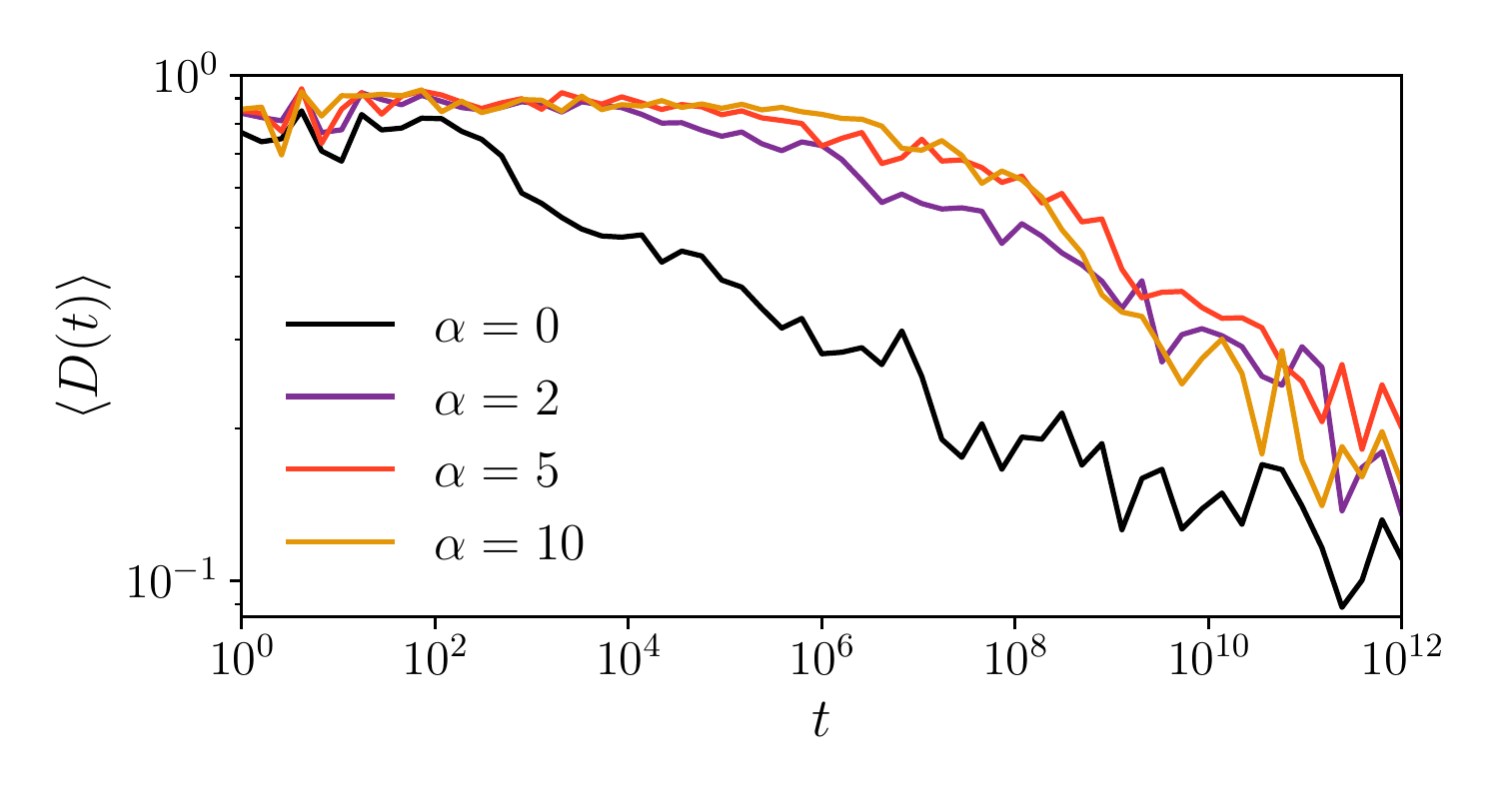}
    \caption{Upper panel: the decay of the normalized QMI, $\mathcal{I}_x / \mathcal{I}_1$, plotted as a function of distance for a thermalizing system, a Stark MBL system, and system with a purely linear potential. The results correspond to a system with $L=18$, $\Delta=0.1$, and have been averaged over 100 eigenstates from the middle of the spectrum. The 3 sites at both ends of the chain have been excluded to minimize boundary effects. Lower panel: the decay of the DEER response with and without curvature. The parameters used were  $L = 12$, $M = 50$, $\Delta = 0.1$, $\gamma = 3$, $d = 7$ and region II being of size $N = 3$.}
    \label{fig:DEER_QMI_fail}
\end{figure}

\subsection{The effect of curvature on the QMI and the DEER response}

We finish this section by demonstrating how the effects of a purely linear potential appear in the experimental probes discussed above.
The spatial decay of the normalized QMI is shown in the upper panel of Fig.~\ref{fig:DEER_QMI_fail}. The results correspond to a system with $L=18$ and have been averaged first over all combinations of sites and then the 100 states closest to the middle of the spectrum. The three sites at the ends of the chain have been excluded from the analysis to minimize boundary effects.
We see that the QMI does not decay with distance in the absence of curvature, but plateaus at large distances, similar to the thermalizing example.
Similarly for the DEER response, we observe in the lower panel of Fig.~\ref{fig:DEER_QMI_fail} that the absence of curvature leads to a much earlier decay compared to the case in which it is present. We furthermore see that the response does not vary significantly with curvature strength, and the smallest $\alpha$ we consider is sufficient to localize the system (as seen in Ref.~\onlinecite{Schulz2019Stark}).
We emphasize that these arguments and observations are made in the setting of finite-sized systems, which is relevant to experiment, but extrapolating to the thermodynamic limit is non-trivial due to the unbounded potential.

\section{Conclusion and perspectives}

In this paper we have explored a number of experimentally relevant probes for conventional MBL in the context of the recently discovered Stark MBL.
We have shown that the outcomes of these experiments are similar to those found in a conventional MBL setting, and as such that they are viable probes for Stark MBL.
These results emphasize the fact that the phenomenology of a Stark MBL system is almost identical to that found in conventional MBL systems.
Other potential probes of Stark MBL include the quantum Fisher information \cite{Smith2016Many-body} and configurational correlators \cite{Lukin2019Probing}, which have been employed in recent experiments studying conventional MBL.
We have also found evidence for Stark MBL in the experimentally relevant setting of interacting bosons in an almost linear potential.

Lastly we have analyzed special features of a purely linear potential, and their
fate under the addition of small amounts of curvature (or disorder), pointing out connections to fractonic systems with conserved dipole moment and Hilbert space fragmentation. 
Indeed, this set-up may very well turn out to be particularly suitable for experimentally probing the physics of Hilbert space fragmentation and fractons in a tunable way.

During the preparation of this manuscript, we have become aware of several other works that explore the links between Hilbert space fragmentation and Stark MBL \cite{Khemani2019Localization,Moudgalya2019Thermalization}.\\

\noindent \textit{Acknowledgements} - The authors are grateful for insightful conversations with Vedika Khemani, Evert van Nieuwenburg, Abhinav Prem, Michael Pretko, and Pablo Sala.
MS is supported by a Google Faculty Award.
FP acknowledges the support of the DFG Research Unit FOR 1807 through grants no. PO 1370/2-1, TRR80, the Nanosystems Initiative Munich (NIM) by the German Excellence Initiative, the Deutsche Forschungsgemeinschaft (DFG, German Research Foundation) under Germany's Excellence Strategy – EXC-2111-390814868 and the European Research Council (ERC) under the European Union's Horizon 2020 research and innovation program (grant agreement no. 771537). RM likewise thanks DFG for support under grant ct.qmat (EXC 2147, project-id 39085490).

\bibliography{MBLbib}

\appendix

\section{Late-time DEER response}
\label{app:DEER}

\begin{figure}
    \centering
    \includegraphics[width=\columnwidth]{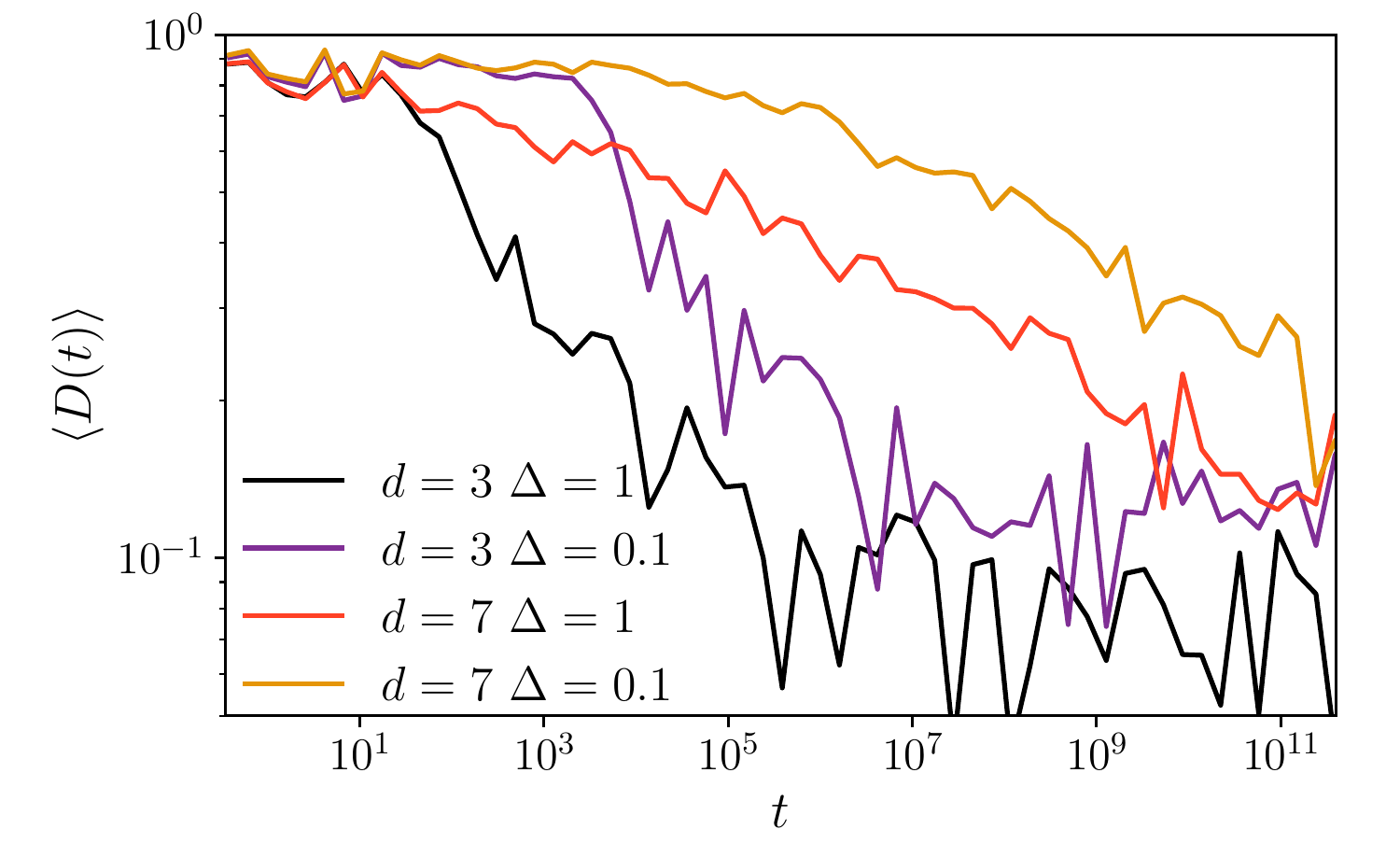}
    \caption{The power-law decay of the DEER response at long times on a double-logarithmic scale.
    The results correspond to a system with $L = 12$, $M = 50$, $\gamma = 3$, $\alpha=2$, and with region II being of size $N = 3$.}
    \label{fig:D_t_PL}
\end{figure}

Besides the short- to medium-time response of the DEER protocol, we have checked whether we can observe the long-time slow decay of the DEER response, as predicted in Ref.~\onlinecite{Serbyn2014Interferometric}. It was shown that upon disorder averaging (in the conventional MBL regime) the DEER response can be approximated by the following equation:
\begin{equation}
    \mathcal{D}(t) =\begin{cases}
    \left(1 + t^2/t^{2}_{0}\right)^{-\beta/2} & t \lesssim t_0 e^{N/\xi}\\
    2^{-N} & t \gg t_0 e^{N/\xi},
  \end{cases}
  \label{eq:DEER_prediction}
\end{equation}
where $t_0 \equiv \hbar/\mathcal{J}_{\text{I$k$}}$ ($k = \text{I} + d$ is the spin in region II that is most strongly coupled to I), and $\beta = \xi \ln 2$ \cite{Serbyn2014Interferometric}. Fig.~\ref{fig:D_t_PL} shows a power-law decay over several decades as well as the saturation at long times (on a double logarithmic scale). Again, the results are less clean than for conventional MBL as we cannot average over realizations of disorder. The behavior is qualitatively similar to that for a conventional MBL system with uncorrelated on-site disorder. The results are obtained for weak ($\Delta = 0.1$) and moderate ($\Delta = 1$) interaction strengths, and both small ($d = 3$) and large ($d = 7$) separations between spin I and region II. The slope of the linear potential in this case is $\gamma = 3$, with a curvature $\alpha = 2$.
We note that the non-interacting Wannier-Stark eigenstates are localized more strongly than exponentially. As a result, if it were possible to study a very large system (with a very large region II), the asymptotic decay would in fact be slower than a power-law according to the assumptions in Ref.~\onlinecite{Serbyn2014Interferometric}.

\begin{figure}
    \centering
    \includegraphics[width=\columnwidth]{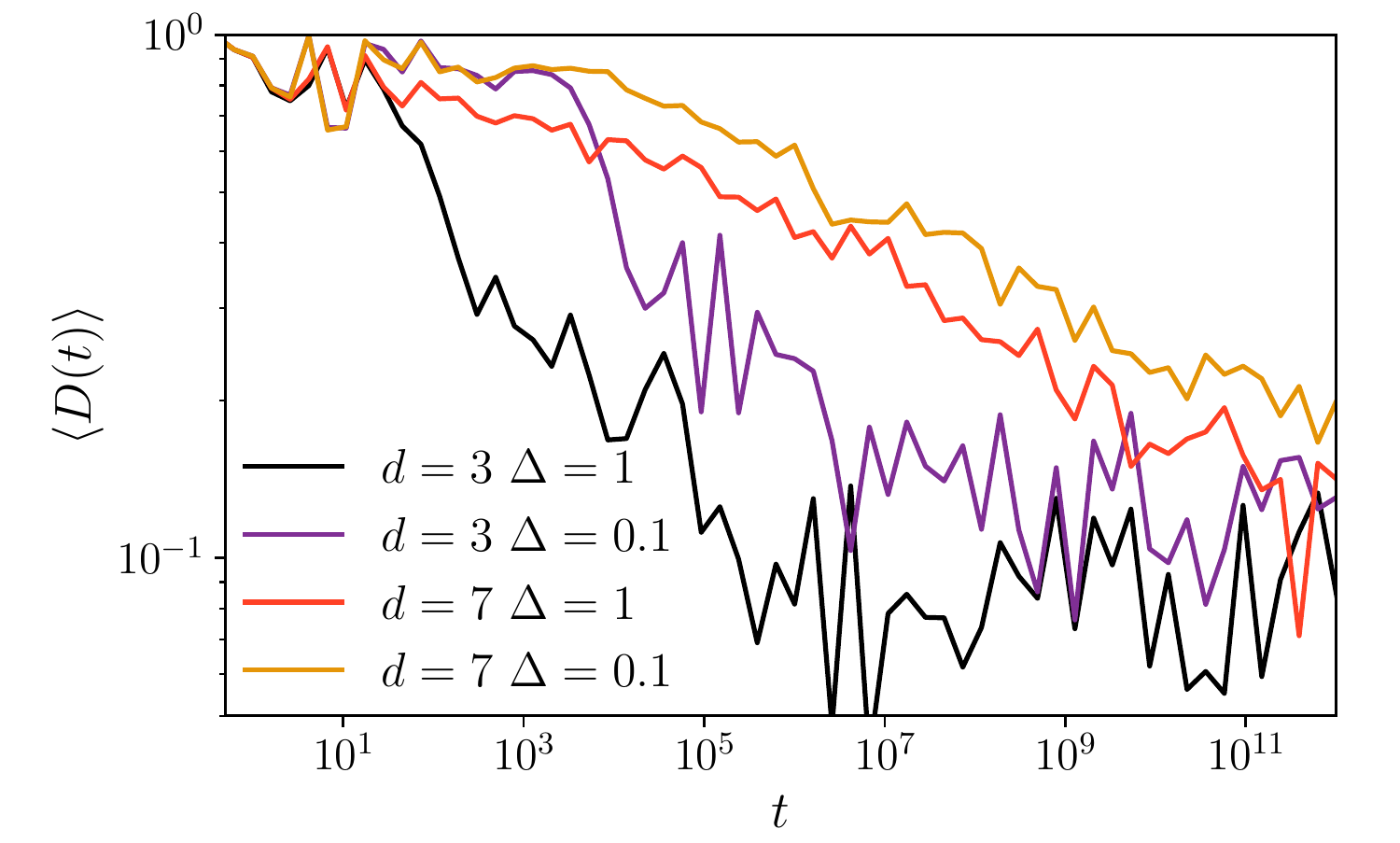}
    \caption{The decay of the DEER response when simulating product initial states in the physical spin basis.
    The decay is qualitatively similar to that seen in Fig.~\ref{fig:D_t_PL}.
    The parameters used were  $L = 12$, $M = 50$, $\gamma = 3$, $\alpha = 2$ and with region II being of size $N = 3$.}
    \label{fig:D_t_PS}
\end{figure}

We have also verified that a similar DEER response is observed when the initial state is a product state in the physical spin basis.
The result is averaged over 50 randomly chosen product states, denoted similarly by the angle brackets. This is shown in Fig.~\ref{fig:D_t_PS}, where we see a similar power-law decay.

\begin{figure}
    \centering
    \includegraphics[width=\columnwidth]{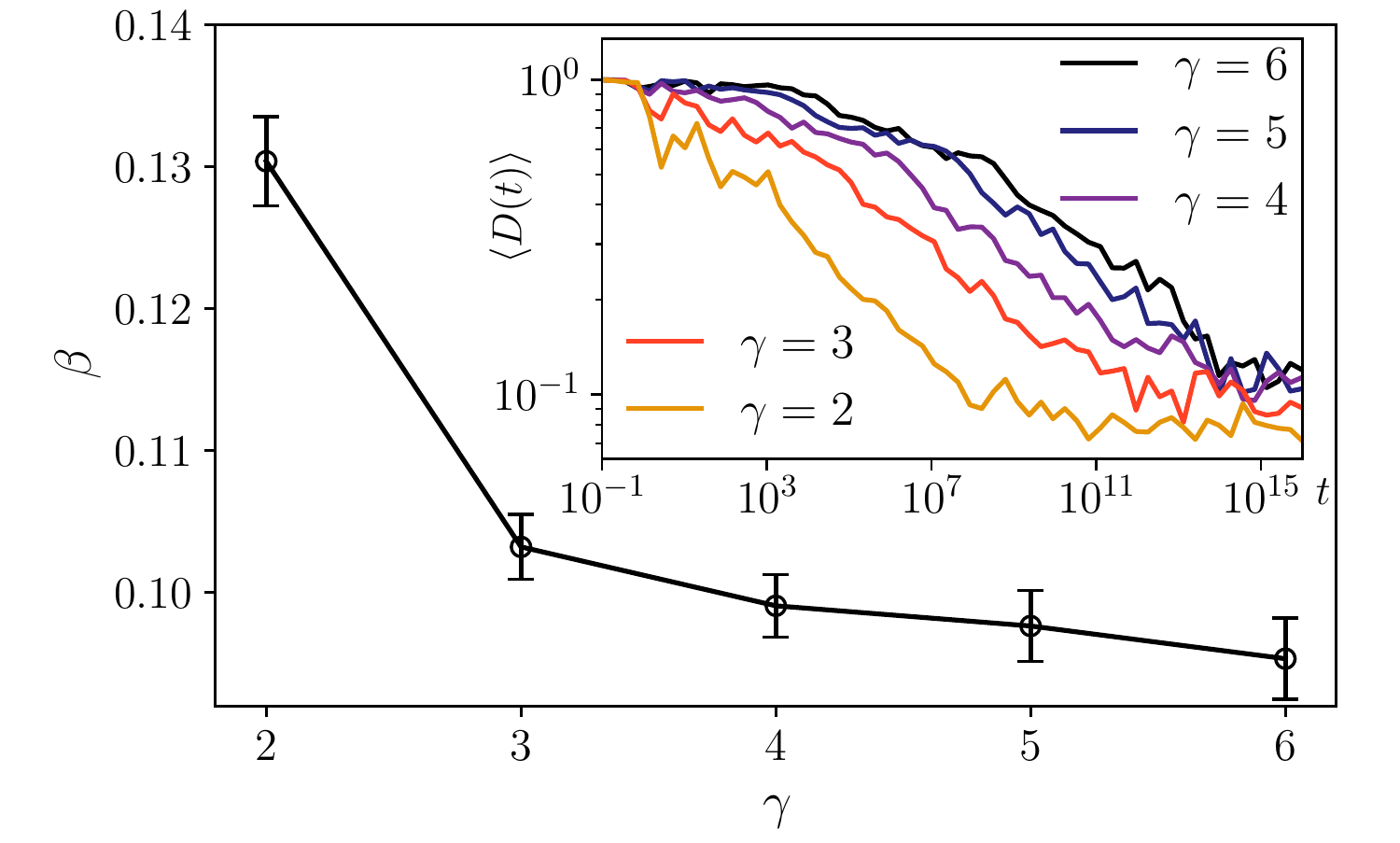}
    \caption{The decay exponent of the long-time DEER response $\beta$ as a function of the linear potential strength $\gamma$. The fit was obtained from the curves in the inset, which correspond to $L = 12$, $M = 50$, $\alpha = 2$, $d=7$, $\Delta = 1$ and $N = 3$.}
    \label{fig:beta_vs_gamma}
\end{figure}

The exponent $\beta$, which describes the power-law decay of the DEER response, is proportional to the localization length $\xi$, and should therefore decrease as the strength of the linear potential is increased.
We confirm this behavior in Fig.~\ref{fig:beta_vs_gamma}, where we show the $\gamma$-dependence of $\beta$, determined by fits to the numerical data (shown in the inset).


\section{Numerical results on the two-particle system}
\label{app:ED2P}

\begin{figure}[h!]
    \centering
    \includegraphics[width=\columnwidth]{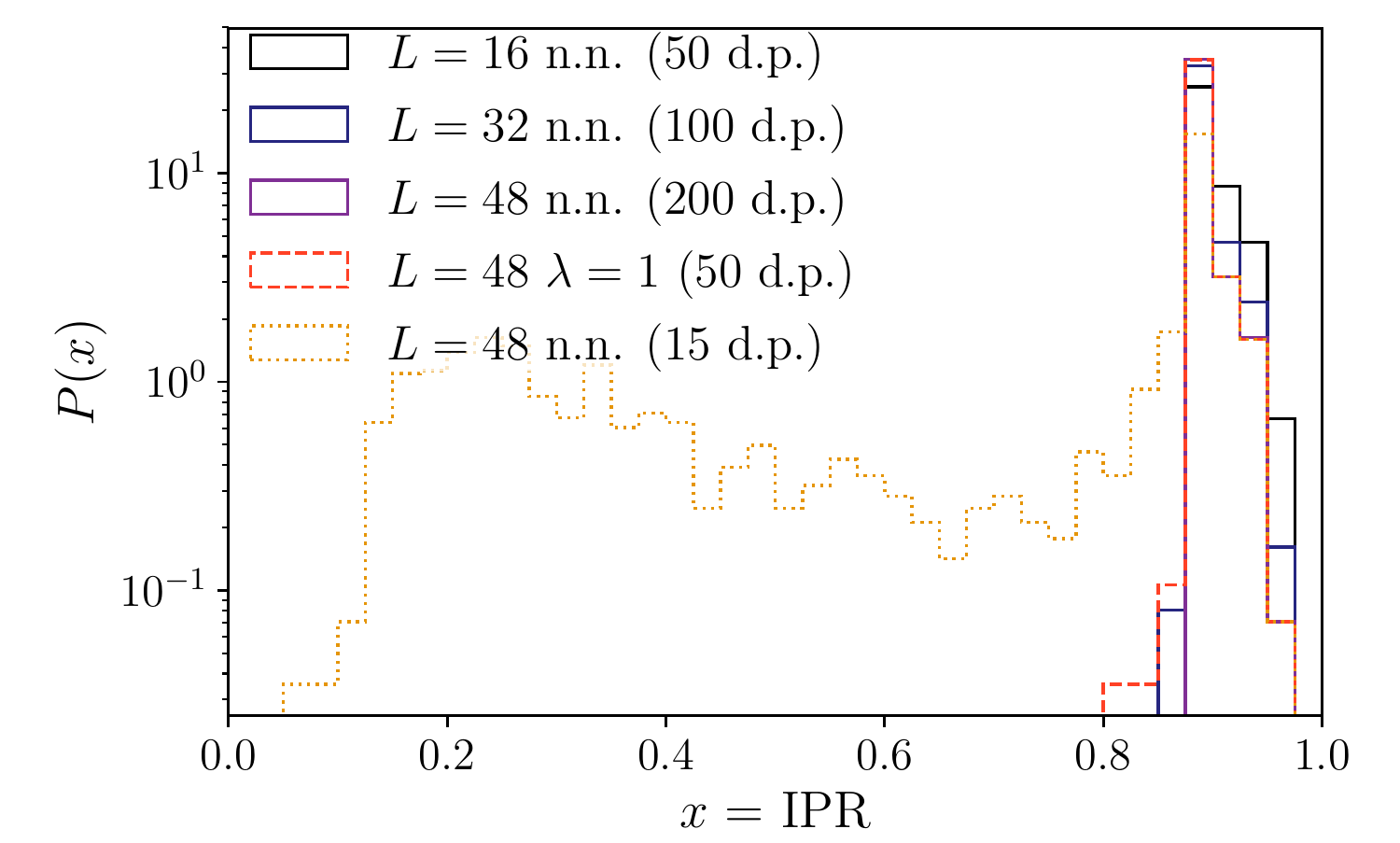}
    \caption{Histograms of the inverse participation ratio in the two-particle system, calculated for several system sizes $L$ for nearest-neighbor (n.n.) and exponentially decaying interactions with high numerical precision (indicated in the legend). The same histogram for $L=48$ and nearest-neighbor interactions calculated with standard precision is shown for comparison (dotted histogram).}
    \label{fig:IPRs}
\end{figure}

Here we present some numerical results on the two-particle system described in Section~\ref{sec:twoparticles}, showing that we see no signatures of delocalization in a purely linear potential even in the presence of interactions.
We can study this system at much larger $L$ than the many-particle problem because the state space grows only quadratically with system size: $\mathcal{N}_2 = L (L-1) / 2$, but the requirements on numerical precision prevent us from going beyond $L = 48$.
We fully diagonalize the Hamiltonian numerically and use the eigenstate inverse participation ratio (IPR) as our measure of delocalization:
\begin{equation}
\mathrm{IPR}_q = \sum_{m,s} \left| \langle m;s | q \rangle \right|^4,
\end{equation}
where $| q \rangle$ is an eigenstate and the sum runs over all $|m; s \rangle$ states (states labelled by the particles' centre of mass $m$ and separation $s$).
The typical IPR decreases with system size for delocalized states but is independent of the system size for a localized state.
The results shown correspond to particles with $\gamma=4$ and nearest-neighbor interaction of strength of 0.1.
We also consider exponentially decaying interactions $V(s) = V_0 \exp \left( [1 - |s|] / \lambda \right)$, where $\lambda$ is the range, and the 1 in the exponential sets interaction strength of particles on neighboring sites to $V_0$ (nearest-neighbor interactions correspond to the limit $\lambda \to 0$).
We use the mpmath package \cite{mpmath} to perform the numerical diagonalization, and in the case of nearest-neighbor interactions we find that: for $L = 16$ 50 decimal places of precision is sufficient, while for $L = 32$ we require 100 decimal places, and for $L=48$ we require more than 150 decimal places (systems with longer-ranged interactions require less precision).
Fig.~\ref{fig:IPRs} shows the distribution of the IPR for a variety of system sizes and both exponentially decaying and nearest-neighbor interactions; the IPR is clearly close to 1 for each set of parameters, indicating localization.
The histogram of IPRs calculated at standard precision for $L=48$ and nearest-neighbor interactions is also shown in Fig.~\ref{fig:IPRs} for comparison (the dotted histogram), demonstrating how insufficient precision can result in false signatures of delocalization.

\end{document}